\documentclass[letterpaper, 10 pt, conference]{ieeeconf}  

\IEEEoverridecommandlockouts                              

\usepackage{amsthm}

\usepackage{amsmath, amssymb}
\usepackage{graphicx, wrapfig, url, xcolor, cite, etoolbox}
\usepackage{hyperref}
\usepackage{verbatim}
\usepackage{bm} 
\usepackage{color} 
\usepackage{graphicx} 

\newtheorem{remark}{Remark}
\newtheorem{definition}{Definition}
\usepackage{booktabs, caption, makecell}
\usepackage{threeparttable}
\usepackage{multirow}
\usepackage{tikz}
\usetikzlibrary{positioning}
\usepackage{wrapfig}
\usepackage{multirow}
\usepackage[figuresright]{rotating}

\usepackage{tablefootnote}

\DeclareMathOperator*{\minimize}{minimize}

\DeclareMathOperator*{\subt}{subject \ to}

\usepackage{tikz-network}
\usepackage{listings}
\usepackage{siunitx}
\usepackage{algorithm, algpseudocode}
\usepackage{pgfgantt}
\hypersetup{nolinks=true}

\newcommand{\bx}{\mathbf{x}}

\newcommand{\bTheta}{\boldsymbol{\Theta}}

\newcommand{\real}{\mathbb{R}}

\usepackage{amsthm}

\title{\LARGE \bf Learning interpretable and stable dynamical models via mixed-integer Lyapunov-constrained optimization*
}

\author{Zhe Li$^{1}$ and Ilias Mitrai$^{2}$
\thanks{*IM would like to acknowledge support from the McKetta Department of Chemical Engineering}
\thanks{$^{1}$Zhe Li is with the Department of Chemical Engineering and Materials Science, University of Minnesota, Minneapolis, MN, USA
        {\tt\small li003679@umn.edu}}%
\thanks{$^{2}$Ilias Mitrai is with the McKetta Department of Chemical Engineering, The University of Texas at Austin, Austin, TX, USA
        {\tt\small imitrai@che.utexas.edu}}%
}

\begin{document}

\maketitle
\thispagestyle{empty}
\pagestyle{empty}

\begin{abstract}
In this paper, we consider the data-driven discovery of stable dynamical models with a single equilibrium. The proposed approach uses a basis-function parameterization of the differential equations and the associated Lyapunov function. This modeling approach enables the discovery of both the dynamical model and a Lyapunov function in an interpretable form. The Lyapunov conditions for stability are enforced as constraints on the training data. The resulting learning task is a mixed-integer quadratically constrained optimization problem that can be solved to optimality using current state-of-the-art global optimization solvers. Application to two case studies shows that the proposed approach can discover the true model of the system and the associated Lyapunov function. Moreover, in the presence of noise, the model learned with the proposed approach achieves higher predictive accuracy than models learned with baselines that do not consider Lyapunov-related constraints.

\end{abstract}


\section{INTRODUCTION}

Mathematical models are the foundation of model-based and data-driven control strategies, capturing the dynamic behavior of the system and control-relevant properties, such as dissipativity and stability \cite{khalil2002nonlinear}. Based on the available information regarding the dynamical system, different approaches have been proposed to discover dynamical models from data. For example, if the model structure is known, then only the model parameters must be identified via parameter estimation or system identification \cite{ljung1998system}. 

On the other hand, if the model structure is unknown or partially known, one must search over a space of functions. An approach to achieve this is to parameterize the dynamic model with neural networks, such as recurrent neural networks \cite{wu2020process}, and input-convex neural networks \cite{amos2017input}. Although such deep learning architectures can approximate smooth functions owing to their universal approximation property, they are inherently black box. An alternative is to search for the symbolic form of the differential equations by representing each differential equation as an expression tree \cite{ly2012learning}. Despite the ability of symbolic approaches to search for a function without any a priori assumptions regarding the functional form, the search is expensive and combinatorial. This has motivated the use of basis functions, i.e., the model is an affine combination of nonlinear basis functions \cite{brunton2016sindy}. 

Usually, the objective of the learning task is the prediction error between the data and the model's prediction. For a dynamical system described by ordinary differential equations, $\dot{x}=f(x)$, the model $f$ can be learned by solving an unconstrained optimization problem $\min_{f} \mathbb{E}_{x\sim D} \big[ |\dot{x} - f(x)|\big]$. Although this approach can lead to highly accurate models, the learned model is not guaranteed to have the same dynamical properties as the original system. For example, given noisy data, the learned model can be highly accurate on the training and testing sets but unstable when evaluated across the entire domain, i.e., a trajectory starting from some initial condition does not converge to the equilibrium. The standard approach to overcome this limitation is a post-learning analysis, where nonlinear systems theory tools, such as local linearization and Lyapunov stability analysis, are used to verify the local/global stability of the learned model. 

An alternative is to incorporate stability requirements into the learning task and solve a constrained problem 
\begin{equation}
\begin{aligned}
    \min_{f} \ \ & \mathbb{E}_{x\sim D} \big[ |\dot{x} - f(x)|\big] \ \ \text{s.t.} \ \ \text{f is stable}.
\end{aligned}
\end{equation}
In this approach, models that are highly accurate for the given data but unstable are infeasible. Depending on the system, different stability constraints can be used. For example, one can first obtain a model with good accuracy and then update the parameters to satisfy stability constraints \cite{xu2023learning}. An alternative is to use contraction theory \cite{blocher2017learning, tobenkin2010convex, singh2021learning} and energy-preserving properties of quadratic systems \cite{kaptanoglu2021promoting}.

In this paper, we focus on approaches that use the Lyapunov conditions as constraints \cite{khalil2002nonlinear}. In such cases, the learning task is a semi-infinite functional optimization problem presented below
\begin{equation}
    \begin{aligned}
        \min_{f, V} \ \ & \mathbb{E}_{x\sim \mathcal{D}} \big[ |\dot{x} - f(x)|\big]\\
        \text{s.t.} \ \ & V(x) \geq 0 \ \ \forall \ x \in \mathcal{D}\setminus \{0\}\\
        & \dot{V}(x) = \frac{\partial V}{\partial x}^\top f(x) \leq 0 \ \ \forall x \in \mathcal{D}\setminus \{0\}\\
        & V(0)=0,
    \end{aligned}
\end{equation}
where $\mathcal{D}$ is the domain of the variables and assuming that $x=0$ is the equilibrium and asymptotic stability. To improve tractability, sampling techniques are used to obtain a finite-dimensional problem, i.e., satisfy the Lyapunov conditions only at the training trajectories. The solution to this learning task presents two challenges: 1) modeling the differential equations and Lyapunov function, and 2) solving the constrained learning problem. For example, one can postulate a functional form for the Lyapunov function, e.g., quadratic, and then train a model while satisfying stability conditions \cite{boots2007constraint, khansari2011learning, shah2025lyapunov}. An alternative is to use neural networks to parameterize the differential equations and Lyapunov function \cite{kolter2019learning, lawrence2020almost, schlaginhaufen2021learning, yang2022input, zhou2022neural, jena2025lilad} and train them jointly, and convex parameterization of linear models \cite{umenberger2018maximum}. Although the neural network-based approaches can discover stable dynamical models, they are inherently black-box. Verifying these models is challenging since the true dynamical model is unknown and neural network verification is computationally expensive \cite{liu2021algorithms}.   

In this paper, we consider learning interpretable and stable dynamical models. Specifically, we parameterize both the differential equations and the Lyapunov function using basis functions and impose the Lyapunov stability conditions as constraints at the training trajectories. The resulting problem is a nonconvex mixed-integer quadratically constrained optimization problem that can be solved to global optimality. The main advantage of this approach is that the model and the candidate Lyapunov function are obtained in an interpretable form. Applications to two case studies, the damped pendulum and the cross-coupled oscillator, demonstrate the approach's ability to discover the model and its associated Lyapunov function. Moreover, the models identified with the proposed approach have better predictive accuracy under measurement noise compared to models obtained using existing (baselines) interpretable sparse regression methods.  The result of the document is organized as follows: In Section~\ref{sec: proposed approach}, we present the proposed approach, and in Section~\ref{sec: experiments} we present the computational experiments.

\section{Lyapunov constrained model discovery} \label{sec: proposed approach}
We consider an autonomous dynamical system described by a system of differential equations
\begin{equation}
    \begin{aligned}\label{eq:ode}
        \dot{x}_{i} & = f_{i}(\textbf{x}) \ \ \forall \ i=1,...,N_{x},
    \end{aligned}
\end{equation}
where $x_{i}$ is state $i$, \textbf{x} is a vector with the state variables, $N_{x}$ is the number of state variables, $\mathcal{N}_{x}$ is the set of state variables, and $f_{i}: \real^{{N}_{x}} \mapsto \real$ is the $i^{\rm{th}}$ differential equation. We assume that a set of trajectories is available and define $\mathcal{J}$ as the index set of the training data and $D$ as the training data $D = \{x_{ij}, \dot{x}_{ij}\}_{i\in \mathcal{N}_{x}, j \in \mathcal{J}}$. 

\subsection{Modeling the differential equations}
To model the differential equations, we use a set of basis functions. We define as $\mathcal{K}_{f}$ the index set of the basis functions used for the dynamic model. Under this parameterization, the differential equation for state $i$ is modeled as follows
\begin{equation} \label{eq: model f generic}
    f_{i}(\textbf{x}) \approx \sum_{k \in \mathcal{K}_{f}} c_{ik} \phi_{k}(\textbf{x}),
\end{equation}
where $c_{ik} \in [c^{\rm{lb}}, c^{\rm{ub}}]$ is the coefficient for the $k$ basis function in the differential equation for variable $i$, and $\phi_{k}$ is the $k^{\rm{th}}$ basis function. For the given training data set, the above equation is written as
\begin{equation}\label{eq: predict dx dt}
    \hat{f}_{ij} = \sum_{k \in \mathcal{K}_{f}} c_{ik} \phi_{k}^{f}(\textbf{x}_{j}) \ \ \forall \ i \in \mathcal{N}_{x}, j \in \mathcal{J},
\end{equation}
where $\hat{f}_{ij}$ is the predicted derivative of variable $i$ at data point $j$, and $\textbf{x}_{j}$ is a vector with the values of the state variables at data point $j$.

We also define a binary variable $z_{ik}^{f}$ which is equal to one if basis function $k$ is used in the expression for the $i^{\rm{th}}$ differential equation and zero otherwise. The binary variables constrain the values of the basis functions coefficients as follows
\begin{equation}
    \begin{aligned}
        & c_{ik} \geq c^{\rm{lb}} z_{ik}^{f} \ \ \forall \ i \in \mathcal{N}_{x}, k \in \mathcal{K}_{f}\\
        & c_{ik} \leq c^{\rm{ub}} z_{ik}^{f} \ \ \forall \ i \in \mathcal{N}_{x}, k \in \mathcal{K}_{f}.
    \end{aligned}
\end{equation}

\subsection{Modeling the Lyapunov function}
Similar to the differential equations, we parameterize the Lyapunov function using a set of basis functions denoted as $\mathcal{K}_{v}$. Under this representation, the Lyapunov function is equal to
\begin{equation}
    V(\textbf{x}) = \sum_{k \in \mathcal{K}_{v}} v_{k} \phi_{k}^{V} (\textbf{x}),
\end{equation}
where $v_{k} \in [v^{\rm{lb}}, v^{\rm{ub}}]$ is the constant for the $k^{\rm{th}}$ Lyapunov basis function $\phi_{k}^{V}$. We define $V_{j} \geq 0$ as the value of the Lyapunov function at data point $j$ and the above equation can be written as
\begin{equation}
    V_{j} = \sum_{k \in \mathcal{K}_{v}} v_{k} \phi_{k}^{V} (\textbf{x}_{j}) \ \ \forall \ j \in \mathcal{J}.
\end{equation}
We also define a binary variable $z_{k}^{V}$ which is equal to one if basis function $k$ is used in the expression for the Lyapunov function and zero otherwise. These binary variables for the Lyapunov's function basis functions constrain the values of the coefficients $v_{k}$ as follows
\begin{equation}
    \begin{aligned}
        & v_{k} \leq v^{\rm{ub}} z_{k}^{V} \ \ \forall \ k \in \mathcal{K}_{V}\\
        & v_{k} \geq v^{\rm{lb}} z_{k}^{V} \ \ \forall \ k \in \mathcal{K}_{V}.
    \end{aligned}
\end{equation}
\subsection{Enforcing Lyapunov conditions}
For a given dynamical model, the existence of a Lyapunov function guarantees stability of the equilibrium point as presented below:
\begin{definition}[Lyapunov function \cite{khalil2002nonlinear}]
Let $x=0$ be an equilibrium point for $\dot{x}=f(x)$, with $f: \mathcal{D} \mapsto R^{n}$. Let $V: \mathcal{D}\mapsto R$ be a continuously differentiable function on a neighborhood $\mathcal{D}$ of $x=0$ such that
\begin{equation}\label{eq:lyap-psd}
    V(0)=0, \ V(x)>0 \ \forall x \in \mathcal{D}\setminus \{0\}
\end{equation}
and
\begin{equation}\label{eq:lyap-deriv}
    \dot V(x)=\nabla V(x)^\top f(x) \le 0 \ \forall x \in \mathcal{D} \setminus  \{0\}.
\end{equation}
then $x=0$ is asymptotically stable. 
\end{definition}
Depending on the properties of the dynamical system, different types of stability can be defined. For example, if the system is exponentially stable, then $\dot{V}(x) \leq -\alpha V(x)$. We define the index set $\bar{\mathcal{J}} = \{ j \in \mathcal{J}: |\textbf{x}_{j}| = 0\}$, which contains the data points that correspond to the equilibrium point, i.e., $|\textbf{x}_{j}|=0$. Given the basis function modeling of the Lyapunov function presented above, the two constraints in Eq.~\ref{eq:lyap-psd} can be written as follows
\begin{equation} \label{eq: V pos per data point}
    \begin{aligned}
        & V_{j} \geq \alpha_{1} |\textbf{x}_{j}| \ \ \forall \ j \in \mathcal{J}\setminus \bar{\mathcal{J}}
    \end{aligned}
\end{equation}
\begin{equation}\label{eq: V zero per data point equil}
    \begin{aligned}
        & V_{j} = 0 \ \ \forall \ j \in \bar{\mathcal{J}},
    \end{aligned}
\end{equation}
where the first constraint (Eq.~\ref{eq: V pos per data point}), with $\alpha_1>0$, enforces the output of the Lyapunov to be positive if $|\textbf{x}| \neq 0$ and the second constraint (Eq.~\ref{eq: V zero per data point equil}) enforces that $V(\textbf{0}) = 0$. We note that in the code implementation of the model, this constraint is relaxed as $V_{j} \leq \delta \ \forall \ j \in \bar{\mathcal{J}}$, with $\delta$ being a small constant.

The time derivative of the Lyapunov function is equal to
\begin{equation*}
    \dot{V}(\textbf{x}) = \sum_{i \in \mathcal{N}_{x}} \frac{\partial V}{\partial x_{i}} (\textbf{x}) f_{i}(\textbf{x}).
\end{equation*}
Under the basis function parameterization, the partial derivative of the Lyapunov function with respect to the state variables is equal to
\begin{equation*}
    \frac{\partial V}{\partial x_{i}} (\textbf{x}) = \sum_{k \in \mathcal{K}_{V}} v_{k} \frac{\partial \phi_{k}^{V}}{\partial x_{i}} (\textbf{x}),
\end{equation*}
where $\partial \phi_{k}^{V}/\partial x_{i}$ is the partial derivative of the $k^{\rm{th}}$ basis function with respect to variable $x_{i}$. We note that the values of these partial derivatives can be computed a priori for each data point. 

The requirement regarding the negative semi-definite nature of the derivative of the Lyapunov function (Eq.~\ref{eq:lyap-deriv}) is enforced via the following constraints
\begin{equation} \label{eq: lyapunov decrease}
\begin{aligned}
    \sum_{i \in \mathcal{N}_{x}} \bigg( \sum_{k' \in \mathcal{K}_{V}} v_{k'} \frac{\partial \phi_{k'}^{V}}{\partial x_{i}} (\textbf{x}_{j})\bigg) \bigg(\sum_{k \in \mathcal{K}_{f}} c_{ik} \phi_{k}(\textbf{x}_{j}) \bigg) \leq & \\
 -\alpha_{2} \sum_{k \in \mathcal{K}_{v}} v_{k} \phi_{k}^{V} (\textbf{x}_{j})& ,
    \end{aligned}
\end{equation}
where $\alpha_{2}\geq 0$ is the convergence rate. This constant can either be fixed a priori or can be a variable. In the latter case, this results in a quadratic constraint. This constraint is enforced for all data points that do not correspond to the equilibrium, i.e., $\forall j \in \mathcal{J}\setminus \bar{\mathcal{J}}$. For the data points which correspond to the equilbrium, $|\textbf{x}|=0$ this constraint is written as follows
\begin{equation} \label{eq: lyapunov decrease at equilibirum}
    \sum_{i \in \mathcal{N}_{x}} \bigg( \sum_{k' \in \mathcal{K}_{V}} v_{k'} \frac{\partial \phi_{k'}^{V}}{\partial x_{i}} (\textbf{0})\bigg) \bigg(\sum_{k \in \mathcal{K}_{f}} c_{ik} \phi_{k}(\textbf{0}) \bigg) = 0.
\end{equation}
The constraints in Eq.~\ref{eq: lyapunov decrease} and \ref{eq: lyapunov decrease at equilibirum} are nonconvex due to the bilinear terms $v_{k'} c_{ik}$.

\subsection{Complexity of the differential equations and Lyapunov function}
The incorporation of binary variables for selecting a basis function for the differential equation and the Lyapunov function enables direct control over the maximum model complexity. This can be done using the following constraints
\begin{equation}
    \sum_{i \in \mathcal{N}_{x}} \sum_{k \in \mathcal{K}_{f}} z_{ik}^{f} \leq \mathcal{C}_{f}
\end{equation}
\begin{equation}
    \sum_{k \in \mathcal{K}_{V}} z_{k}^{V} \leq \mathcal{C}_{V},
\end{equation}
where $\mathcal{C}_{f}$ is the maximum number of basis functions that can be used for all differential equations and $\mathcal{C}_{V}$ is the maximum number of basis functions that can be used for the Lyapunov function. 

\subsection{Overall training problem}
The objective of the learning problem has three components: the prediction error $\mathcal{L}_{a}$, the complexity of the differential equations $\mathcal{L}_{c}^{f}$ and the Lyapunov function $\mathcal{L}_{c}^{V}$. These terms are computed as follows
\begin{equation}
\begin{aligned}
        \mathcal{L}_{a} & = \frac{1}{N_{D}} \sum_{i \in \mathcal{N}_{x}}\sum_{j \in \mathcal{J}} |\dot{x}_{ij} - \hat{f}_{ij}|\\
        \mathcal{L}_{c}^{f} & = \sum_{i \in \mathcal{N}_{x}} \sum_{k \in \mathcal{K}_{f}} z_{ik}^{f}\\
        \mathcal{L}_{c}^{V} & = \sum_{k \in \mathcal{K}_{V}} z_{k}^{V}.
\end{aligned}
\end{equation}
The absolute values can be reformulated using linear constraints. Specifically, we define nonnegative variables $\epsilon_{ij}^{+}$, $\epsilon_{ij}^{-}$, and reformulate each absolute value as follows
\begin{equation}
\dot{x}_{ij} - f_{ij} = \epsilon_{ij}^{+}-\epsilon_{ij}^{-} \ \ \forall \ i \in \mathcal{N}_{x}, j \in \mathcal{J}.
\end{equation}
The prediction over all data points can be written as
\begin{equation}
\begin{aligned}
        \mathcal{L}_{a} & = \frac{1}{N_{D}} \sum_{i \in \mathcal{N}_{x}}\sum_{j \in \mathcal{J}} \big(\epsilon_{ij}^{+} + \epsilon_{ij}^{-}\big).
\end{aligned}
\end{equation}
Overall, the learning task is 
\begin{equation}\label{eq: monolithic}
  \begin{aligned}
  \minimize_{c_{ik}, v_{k},z^f_{ik},z^V_{k}} \quad
    & \mathcal{L}_a 
  + \omega_1 \mathcal{L}_c^f
  + \omega_2 \mathcal{L}_c^V \\
  \subt \quad
    & \text{Eq.}~5,6,8,9,12-17\\
  \end{aligned}
\end{equation}
This is a nonconvex quadratically constrained mixed integer optimization problem due to constraints in Eq.~\ref{eq: lyapunov decrease} and \ref{eq: lyapunov decrease at equilibirum}. 

\begin{remark}
\normalfont The solution to this learning problem is not guaranteed to yield a stable dynamical model over the entire domain. Depending on the objective used and the amount of data, the optimization problem can be degenerate, since, for a given differential equation, multiple Lyapunov functions can yield the same optimal objective value. This limitation is due to the finite amount of data used for training. Moreover, branch-and-bound solvers stop the search once a primal solution that satisfies a predetermined optimality-gap tolerance is found. Hence, although a valid Lyapunov function can be in the feasible space, it might not be bound due to the branching policy of the solver.
\end{remark}

\begin{remark}
    \normalfont The learned differential equations and Lyapunov function can be verified using standard approaches, i.e., checking if 
    \begin{equation*}
        \begin{aligned}
            & V^{\rm{min}} \in \arg \min_{x \in \mathcal{D}\setminus \{0\}} V(x) \rightarrow V^{\rm{min}} \geq 0\\
            & \dot{V}^{\rm{max}} \in \arg \max_{x \in \mathcal{D}\setminus \{0\}} \dot{V}(x) \rightarrow \dot{V}^{\rm{max}} \leq 0.
        \end{aligned}
    \end{equation*}
Since the Lyapunov function and the model are in algebraic form, existing state-of-the-art global optimization algorithms can be used to compute $V^{\rm{min}}$ and $\dot{V}^{\rm{max}}$. However, because the identified model is not guaranteed to be the true model, the resulting Lyapunov function is not necessarily valid for the system, as the true model is unknown. We note that several approaches use neural network verification approaches to certify a Lyapunov function \cite{ravanbakhsh2019learning}. However, in such cases, either the true model is known or the Lyapunov function is parameterized in such a way that it always satisfies the Lyapunov conditions. 
\end{remark}

\begin{remark}
    \normalfont In cases where the returned candidate Lyapunov function is not valid over the entire domain, several approaches can be followed to search for a valid Lyapunov function. An approach is to generate more training data and resolve the learning task, which is commonly in existing methods for learning neural network Lyapunov functions \cite{ravanbakhsh2019learning}. An alternative is to add integer cuts and resolve the learning task. This approach does not require generating extra training data. Finally, one could exclude the current values of the Lyapunov function constants using a trust region-like approaches.
\end{remark}

\section{Computational Experiments} \label{sec: experiments}
In this section, use the proposed approach to discover stable dynamical models for two case studies. As presented above, the learning task is a nonconvex mixed-integer quadratically constrained program (MIQCP) solved to global optimality using Gurobi \cite{gurobi} (Version 12.0.3). 

We consider two evaluation metrics: vector field error and normalized coefficient error. Let $\textbf{c}^{*}$ be the coefficients of the learned dynamics and $\textbf{c}^{true}$ be the coefficients of the ground truth dynamics. The vector field error is defined as the pointwise $\ell_2$ error of the derivatives $\| \bTheta(\bx) \textbf{c}^* - \bTheta(\bx) \textbf{c}^{true}\|_2$, where $\bTheta(\bx)$ are the basis functions. To compute this error, we discretize the state space and evaluate the vector field error at each grid point. The relative coefficient error is defined as the normalized $\ell_2$ error between the learned and true coefficients $\| \textbf{c}^* - \textbf{c}^{true}\|_2 / | \textbf{c}^{true}\|_2$. The source code for the experiments is available on GitHub: \href{https://github.com/BambooSticker/LyapSINDy}{https://github.com/BambooSticker/LyapSINDy}.

\subsection{Case study 1: Damped pendulum}
First, we use the damped pendulum as an illustrative case study to show that the proposed approach can discover the differential equations and the Lyapunov function. The dynamic behavior of the system is described by the following system of ordinary differential equations
\begin{equation}
  \begin{aligned}\label{eq:pen-ode}
    & \dot{x}_1 = x_2 \\
    & \dot{x}_2 = - \sin(x_1) - x_2.
  \end{aligned}
\end{equation}
A Lyapunov function for this system is the total system energy, which is equal to 
\begin{equation}
  V(x_{1},x_{2}) = (1-\cos(x_1)) + \frac{x_2^2}{2}.
\end{equation}

We generate a single trajectory from the initial condition $[-2, -1.5]$ with a sampling interval of $\Delta t=0.05$, resulting in $|\mathcal{J}|=400$ data points. The coefficients in the objective function are equal to $\omega_1=\omega_2=5\times 10^{-3}$, and $c^{\rm{ub}}=1$, $c^{\rm{lb}}=-1$. We set $\alpha_1=0.2$ as a parameter and $\alpha_2\ge1\times 10^{-5}$ as a variable. 
We allow at most five terms in both differential equations, i.e., $\mathcal{C}_{f}=5$, and five terms in the Lyapunov function, i.e., $\mathcal{C}_{V}=5$. 

\begin{figure}[ht]
  \centering
  \includegraphics[scale=0.6]{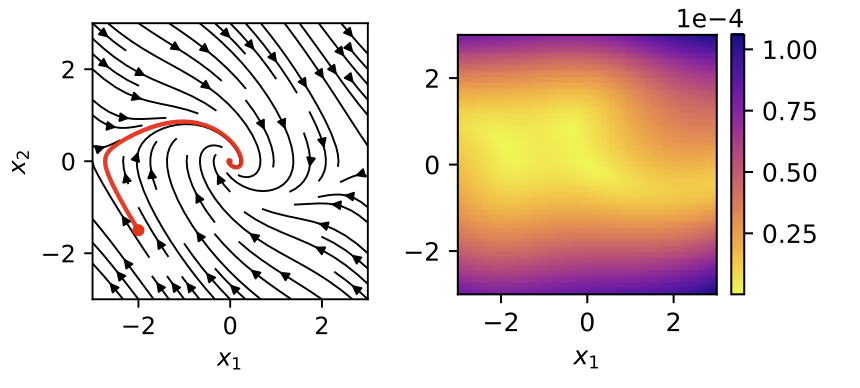}
  \caption{Phase portrait of the damped pendulum system (left) and the $\ell_2$ error of the learned vector field (right). }
  \label{fig:pend-l2}
\end{figure}

The learning task is solved in 2.7 seconds and the learned constants for the basis functions are presented in Table~\ref{tb:pend-coeff}. From the results, we observe that with a single trajectory and without noise, the proposed approach returns the correct coefficients for the basis functions at the differential and the Lyapunov function. The phrase portrait of the ground truth dynamics and the sampled trajectory are presented in the left panel of Fig.~\ref{fig:pend-l2} and in the right panel, the pointwise $\ell_{2}$ error of the vector field is identified. From the results, we observe that the error over the state space is below $1\times10^{-4}$.

We also solve the learning task for different complexities of the Lyapunov function. Specifically, we keep the maximum number of basis functions for the differential equations equal to five $\mathcal{C}_{f}=5$ and allow only one basis function of the Lyapunov function $\mathcal{C}_{V}=1$. In this case, the learning task is declared infeasible because a valid Lyapunov function for the given data using at most one basis function does not exist. If the complexity of the Lyapunov function increases to $\mathcal{C}_{V}=2$, then the identified model is not the true model. 

\vspace{0.5em}

\vspace{0.5em}
\begin{table}[ht]
\centering
\caption{Identified dynamical model and Lyapunov function for the damped pendulum system.}
\resizebox{0.35\textwidth}{!}{%
\begin{tabular}{cccc}
\toprule
\multirow{2}{*}{\textbf{Basis function}} & \multicolumn{3}{c}{\textbf{Coefficient}} \\
\cmidrule(lr){2-4}
& $\dot{x}_1$ & $\dot{x}_2$ & $V(\bx)$ \\
\midrule
$1$        & 0.0   & 0.0    & 1.0  \\
$x_1$        & 0.0   & 0.0    & 0.0    \\
$x_2$        & 1.0 & -1.0 & 0.0    \\
$x^2$      & 0.0   & 0.0    & 0.0    \\
$x_1 x_2$       & 0.0   & 0.0    & 0.0    \\
$x_2^2$      & 0.0   & 0.0    & 0.5  \\
$\sin(x_1)$  & 0.0   & -1.0 & 0.0    \\
$\cos(x_1)$  & 0.0   & 0.0    & -1.0 \\
$\sin(x_2)$  & 0.0   & 0.0    & 0.0    \\
$\cos(x_2)$  & 0.0   & 0.0    & 0.0    \\
\bottomrule
\end{tabular}%
}
\label{tb:pend-coeff}
\end{table}

\begin{figure*}[ht]
    \centering
    \includegraphics[scale=0.7]{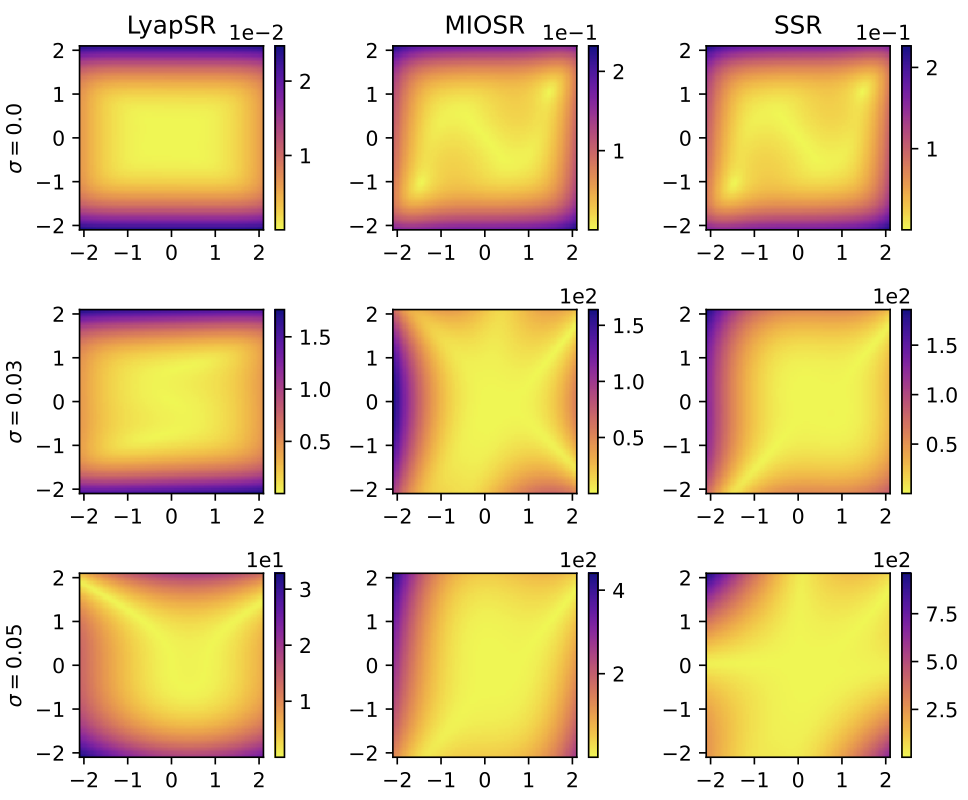}
    \caption{$\ell_2$ vector field error of the learned models produced by all methods under different noise levels. LyapSR refers to the proposed method.}
    \label{fig:cco-l2}
\end{figure*}

\subsection{Case study 2: Cross-coupled oscillator}
We also evaluate the proposed approach in the presence of noise, and compare the accuracy of the learned model with two state-of-the-art sparse regression algorithms: stepwise sparse regression (SSR) \cite{boninsegna2018sparse} and mixed integer optimization-based sparse regression (MIOSR) \cite{bertsimas2023learning}. We note that the MIOSR approach is similar to the proposed method if the Lyapunov conditions (constraints) are not considered. The main challenge in the presence of noise is that a decrease in the Lyapunov function cannot be guaranteed. In this case, the presence of the Lyapunov function does not imply asymptotic (or exponential) stability due to the noise. 

We consider the following cubic cross-coupled oscillator:
\begin{equation}
    \begin{aligned}
        & \dot{x}_1 = -x_1 + x_2 - 2x_1^3\\
        & \dot{x}_2 = - x_2 - 3x_1 - 2x_2^3.
    \end{aligned}
\end{equation}
A candidate Lyapunov function of this system is a simple quadratic function $V(x_1, x_2) = x_1^2 + x_2^2$.

We generate a single trajectory for 6 time units with a sampling interval of $\Delta t=0.01$ from the initial state $[2, 1.5]$, yielding $|\mathcal{J}|=600$ data points. After obtaining the dataset, we add Gaussian noise $w_j=\mathcal{N}(0,\sigma^2)$ to each data point, and compare different methods under four noise levels $\sigma \in [0, 0.03, 0.05, 0.1]$. We choose third-order polynomial basis functions to learn the dynamics for all methods, and fourth-order polynomials to learn the Lyapunov function. The parameters of the learning task are equal to $\omega_1=\omega_2=5\times 10^{-3}$. We set $\alpha_1=0.05$ as a parameter and $\alpha_2\ge1\times 10^{-5}$ as a variable. 
The baselines MIOSR and SSR are implemented through the PySINDy package \cite{Kaptanoglu2022} with default parameters.

We first compare the $\ell_{2}$ vector field error for all methods. From Fig.~\ref{fig:cco-l2} we observe that without noise the error for the proposed methods is in the order of $10^{-2}$ whereas for the MIOSR and SSR in the order of $10^{-1}$. As the noise increases, the error for all methods increases. However, the increase is less dramatic for the proposed method. Specifically, for MIOSR and SSR, as $\sigma$ increased from $0.0$ to $0.03$ the error increased in the order of $10^{2}$, whereas with the proposed method, the error was in the order of $10^{0}$ for $\sigma=0.03$ and $10^{1}$ for $\sigma=0.05$. From these results, we observe that incorporating the Lyapunov condition constraints during training, even with a single trajectory, yields more accurate, interpretable models, achieving up to two orders of magnitude higher accuracy than standard baselines.

\begin{figure}[ht]
    \centering
    \includegraphics[scale=0.32]{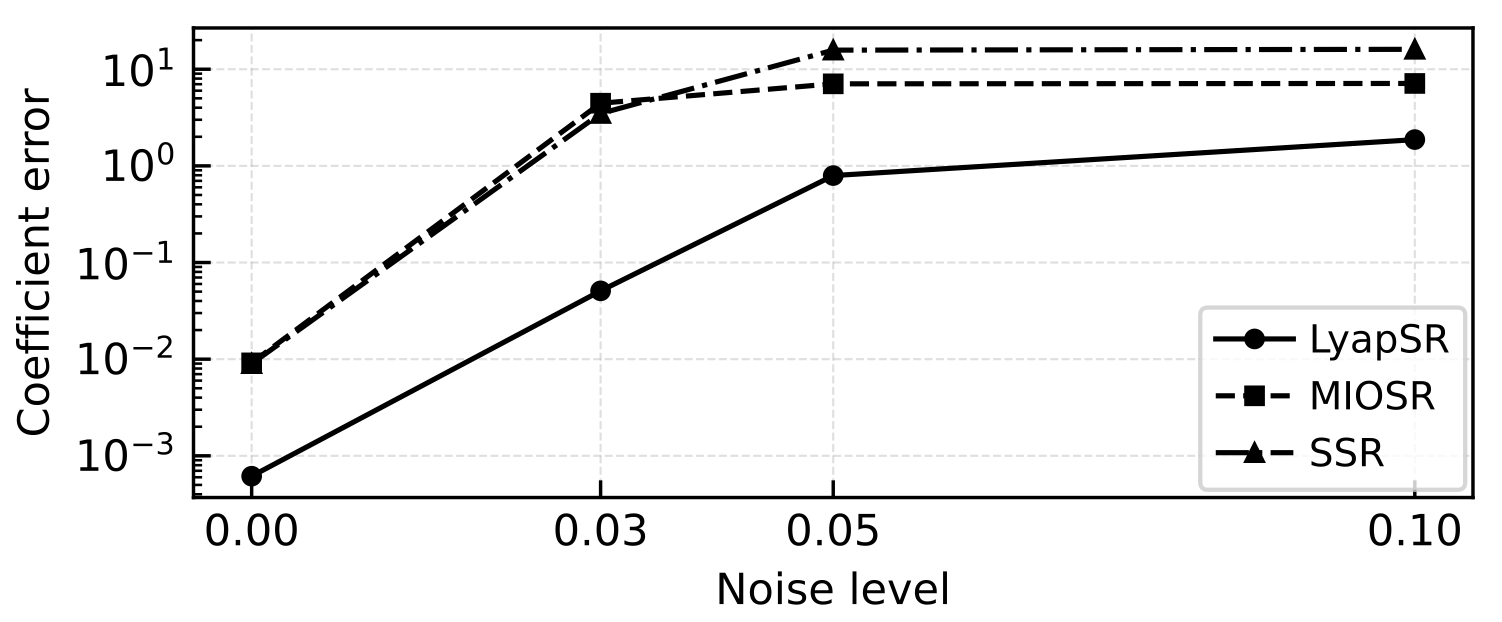}
    \caption{Coefficient error comparison of the learned dynamical models. LyapSR is the proposed approach.}
    \label{fig:cco-coeff}
\end{figure}

Finally, we compare the relative errors of the coefficient values obtained from the different approaches, as shown in Fig.~\ref{fig:cco-coeff}. 
For $\sigma=0$, the proposed approach identifies the Lyapunov function as $V(x_1, x_2)=0.556x_1^2+0.556x_2^2$, which can be proved to be a valid Lyapunov function for the entire domain. In addition, the coefficient error of the proposed approach is on the order of $10^{-3}$, whereas the errors of MIOSR and SSR are on the order of $10^{-2}$. As the noise level increases, the coefficient errors across all methods increase; however, the coefficient error of the proposed approach consistently remains lower. 
For $\sigma=0.03$, the proposed approach correctly identifies the model structure, while the other methods produce models that have different basis functions than the true model. For $\sigma=0.05$, the proposed approach still preserves five of the six correct basis functions. 
These differences also manifest in the $\ell_{2}$ error of the vector field discussed previously. 

\section{Conclusion}
In this paper, we proposed a mixed integer optimization approach for learning interpretable and stable dynamical models from data. The proposed approach uses basis functions to parameterize the differential equations and the Lyapunov functions. The Lyapunov conditions for stability are encoded as constraints; the positive definiteness of the Lyapunov function is a linear constraint, whereas the negative definite derivative is a nonconvex quadratic constraint. The resulting learning task is a mixed integer quadratically constrained optimization problem that can be solved to global optimality using state-of-the-art global optimization solvers. Application to two case studies demonstrates the ability of the proposed approach to identify the dynamical model and the associated Lyapunov function. Moreover, in the presence of noise, the learned model using the proposed approach achieves higher predictive accuracy than existing baseline approaches that do not explicitly enforce Lyapunov conditions. 

Finally, we note that the proposed approach can not guarantee that the returned Lyapunov function is valid from the entire domain, due to the limited data, the lack of access to the true model, and the degeneracy of the learning task, i.e., multiple Lyapunov functions can lead to a dynamical model with the same prediction loss for the training data set. These aspects can be addressed by incorporating more data in the training problem and adding integer cuts to explore other functional forms for the Lyapunov function.


\bibliographystyle{IEEEtran}
\bibliography{reference}

\end{document}